\documentclass[runningheads]{llncs}
\usepackage{graphicx}
\usepackage{amsfonts}
\usepackage{amsmath} 
\usepackage{setspace}
\usepackage{enumerate}
\usepackage[colorlinks, linkcolor=blue, anchorcolor=blue, urlcolor=blue, citecolor=blue]{hyperref}
\usepackage{array}
\usepackage{multirow}
\usepackage{marvosym}

\def\mathbi#1{\textbf{\em #1}}
\begin{document}
\title{A Knowledge Graph based Approach for Mobile Application Recommendation\thanks{Supported by National Key Research and Development Project of China (No. 2019YFB1405302). Please cite the paper as the following:
\textbf{Zhang, M., Zhao, J., Dong, H., Deng, K., Liu, Y.: A Knowledge Graph based Approach for Mobile Application Recommendation. The 18th International Conference on Service Oriented Computing (ICSOC 2020) (2020)}}}
%
%
\author{Mingwei Zhang\inst{1} \and Jiawei Zhao\inst{1} \and Hai Dong\inst{2}\textsuperscript{(\Letter)} \and Ke Deng\inst{2} \and Ying Liu\inst{1}}
\authorrunning{M. Zhang et al.}
\titlerunning{KG based Mobile App Recommendation}
%
\institute{Software College, Northeastern University, Shenyang, China\\
\email{\{zhangmw,liuy\}@swc.neu.edu.cn}, \email{zhaojiawei@stumail.neu.edu.cn}\and
School of Science, RMIT University, Melbourne, Australia\\
\email{\{Hai.Dong,Ke.Deng\}@rmit.edu.au}}

\maketitle              

\begin{abstract} With the rapid prevalence of mobile devices and the dramatic proliferation of mobile applications (apps), app recommendation becomes an emergent task that would benefit both app users and stockholders. How to effectively organize and make full use of rich side information of users and apps is a key challenge to address the sparsity issue for traditional approaches. To meet this challenge, we proposed
a novel end-to-end \underline{K}nowledge \underline{G}raph Convolutional \underline{E}mbedding \underline{P}ropagation Model (KGEP) for app recommendation. Specifically, we first designed a knowledge graph construction method to model the user and app side information, then adopted KG embedding techniques to capture the factual triplet-focused semantics of the side information related to the first-order structure of the KG, and finally proposed a relation-weighted convolutional embedding propagation model to capture the recommendation-focused semantics related to high-order structure of the KG. Extensive experiments conducted on a real-world dataset validate the effectiveness of the proposed approach compared to the state-of-the-art recommendation approaches.

\end{abstract}
\section{Introduction}
Recent years, people have witnessed a rapid prevalence of smart mobile devices and a dramatic proliferation of mobile applications. The large number and high variety of apps are posing a great challenge for users to choose appropriate ones. As a consequence, app recommendation has attracted more and more attention these years. On the one hand, it can help users find their desired or interested apps more easily and quickly. On the other hand, it will benefit developers and stockholders of apps to get more profits in the mobile app ecosystem.

However, sparsity is a typical characteristic of app usage data. For instance, as of May 2020, there are over 2.9 million apps on Google Play\cite{GP}, but most of billions of users only install at most hundreds of apps. To address the sparsity problem of user-app interactions, researchers usually turn to feature-rich scenarios, where side information of users and apps is used to compensate for the sparsity and improve the performance of recommendation. As detailed in section 5, most of them\cite{ARLiu,ARHuang,ARYin,ARZhu,ARCao,ARXu,ARGuo,ARLiang,ARXie} only exploited limited types of side information. In addition, they usually treated different kinds of side information as isolated features of users and apps, and neglected the relations and semantics of them. Consequently, how to effectively organize and make full use of side information of users and apps is a great challenge to make successful app recommendation.

To meet the above challenge, we proposed a KG based app recommendation approach. A KG is a type of directed heterogeneous graph in which nodes correspond to entities and edges correspond to relations\cite{KG}. Among various types of side information, the KG contains much more fruitful facts and introduces semantic relatedness among apps, which can help find their latent connections. Beyond that, the KG consists of relations with various types, which is helpful for exploring a user’s interests reasonably. To be specific, we proposed a KG convolutional embedding propagation model (KGEP) for app recommendation. First, a KG construction method is designed to organize different kinds of side information effectively. Then, a translation based KG embedding model is adopted to capture the general semantics of side information from the perspective of general KG facts. Finally, a relation-weighted KG convolutional embedding propagation model is designed to further capture the recommendation-focused semantics from the perspective of recommendation. We evaluated the proposed model on a real dataset crawled from Google Play. The experimental results verify the effectiveness of our method for app recommendation when compared to the state-of-the-art methods.

The major contributions of this paper are summarized as follows.
\begin{enumerate}
\item It is the first work, to the best of our knowledge, that incorporates a KG to organize and take full advantage of diverse side information for app recommendation.  
\item We proposed a novel end-to-end app recommendation model KGEP, which can capture the semantics of rich side information related to both the first-order and high-order structures of the constructed KG, by utilizing KG general embedding techniques and convolutional propagated embedding techniques respectively.
\item We conducted extensive experiments using a real app dataset. The comparative results demonstrate that our approach achieves higher performance compared to the state-of-the-art recommendation methods.
\end{enumerate}
The remainder of this paper is organized as follows. Sect. 2 formulates the app recommendation problem. Sect. 3 presents the proposed model in detail. Sect. 4 discusses the experimental results. Sect. 5 introduces related works. Finally, we concluded the paper and indicated some future directions in Sect. 6.

\section{Notations and Problem Formulation}
The app recommendation scenario contains a set of users $\mathcal{U} = \{u_1, u_2, …$, $u_{|\mathcal{U}|}\}$, a set of apps  $\mathcal{A} = \{a_1, a_2, …, a_{|\mathcal{A}|}\}$, and their historical interactions. In addition, we have rich side information for users and apps (e.g., app attributes and description texts). Typically, such auxiliary data consists of real-world entities and relationships among them to profile a user or an app. We organized the side information in the form of KG. 

\underline{A}pp \underline{R}ecommendation \underline{K}nowledge \underline{G}raph (\textbf{ARKG}), denoted as $\mathcal{G}$, is a directed graph composed of entity-relation-entity triples $(h, r, t)$, where $h\in \mathcal{E}$, $r\in \mathcal{R}$ and $t\in \mathcal{E}$ are the head, relation, and tail of a knowledge factual triple, and $\mathcal{E}$ and $\mathcal{R}$ are the set of entities and relations in $\mathcal{G}$, respectively. For example, the triple (Facebook, OfferedBy, Facebook) states the fact that the company “Facebook” offers the app “Facebook”. According to the side information which we can crawl and their importance for recommendation, we defined the following 13 types of entities for the ARKG.

\medskip\noindent \textbf{Definition 1 (\textit{Content-Topic} Entity)}. Considering \textit{Readme} texts of apps provided by developers contain rich app profiles and are crucial to the efficacy of the ARKG to do recommendation,  We used probabilistic topic modeling to incorporate them into the ARKG. A \textit{Content-Topic} entity is a distribution over terms, which can be used to explore users’ preference on specific topics. The number of \textit{Content-Topic} entities involved in the ARKG is a hyperparameter and can be configured by recommender service users. 

\medskip\noindent Due to space limitations, the definitions of the other 12 types of entities (i.e., \textit{User}, \textit{App}, \textit{Category}, \textit{Provider}, \textit{Popularity}, \textit{Age-Restriction}, \textit{Ads}, \textit{Fee}, \textit{Interactive-Elements}, \textit{Quality}, \textit{Updated-Time}, \textit{Size}) involved in the ARKG are not listed any more. Based on these kinds of entities, 18 relations were defined for the ARKG, the detailed information of which is listed in Table~\ref{relations}.  

\begin{table}
\caption{Relations involved in the ARKG}\label{relations}
\small
\begin{tabular}{p{0.2\textwidth}|p{0.2\textwidth}|p{0.2\textwidth}|p{0.36\textwidth}}
\hline
Relation & Head entity & Tail entity & Related side information\\
\hline
\textit{INTERACT} & \textit{User} & \textit{App} & User-app interaction data\\
\textit{HAVINGCT} & \textit{App} & \textit{Content-Topic} & Apps' \textit{Readme} texts \\
\textit{HAVINGC} & \textit{App} & \textit{Category} & Apps' category data \\
\textit{OFFEREDBY} & \textit{App} & \textit{Provider} & Apps' provider data \\
\textit{CONTENTR} & \textit{App} & \textit{Age-Restriction} & Apps' content rating data \\
\textit{HAVINGA} & \textit{App} & \textit{Ads} & No ads or not of an app \\
\textit{HAVINGF} & \textit{App} & \textit{Fee} & Free or not of an app \\
\textit{HAVINGIE} & \textit{App} & \textit{Interactive-Elements}  & Apps' interactive-elements data \\
\textit{HAVINGQ} & \textit{App} & \textit{Quality} & Users’ review grades of apps \\
\textit{HAVINGP} & \textit{App} & \textit{Popularity} & Apps' install numbers\\
\textit{HAVINGUT} & \textit{App} & \textit{Updated-Time} & Apps' updated time \\
\textit{HAVINGS} & \textit{App} & \textit{Size} & Apps' size data \\
\textit{USIMILAR} & \textit{User} & \textit{User} & User-app rating matrix \\
\textit{CTSIMILAR}  & \textit{Content-Topic} & \textit{Content-Topic} & \textit{Content-Topic} entity data \\
\textit{QSIMILAR} & \textit{Quality} & \textit{Quality} & \textit{Quality} entity data \\
\textit{PSIMILAR}  & \textit{Popularity} & \textit{Popularity} & \textit{Popularity} entity data \\
\textit{UTSIMILAR}  & \textit{Updated-Time} & \textit{Updated-Time} & \textit{Updated-Time} entity data\\
\textit{SSIMILAR} & \textit{Size} & \textit{Size} & \textit{Size} entity data \\
\hline
\end{tabular}
\end{table}

The relation \textit{INTERACT} denotes historical user-app interactions. The next 11 relations denote that an app has some specific profiles. The last 6 relations denote that one entity is similar to another with the same entity type. Based on the above definitions, how to extract factual triplets and construct the ARKG will be presented in Sect. 3.1. 

We formulated the KG-based app recommendation problem as follows. Given the sets of users and apps, and their side information, we aim to construct an ARKG $\mathcal{G}$. Then taking $\mathcal{G}$ as an input, we aimed to predict whether user $u$ has a potential interest in app $a$ with which she has had no interaction before. Our task can be formulated to learn a prediction function  $\hat{y}_a^u = \mathcal{F}(u,a|\theta,\mathcal{G})$, where $\hat{y}_a^u$ denotes the probability that user $u$ will engage with app $a$, and $\theta$ denotes the model parameters of function $\mathcal{F}$.

\section{Methodology}
\begin{figure}[ht]
\includegraphics[width=\textwidth]{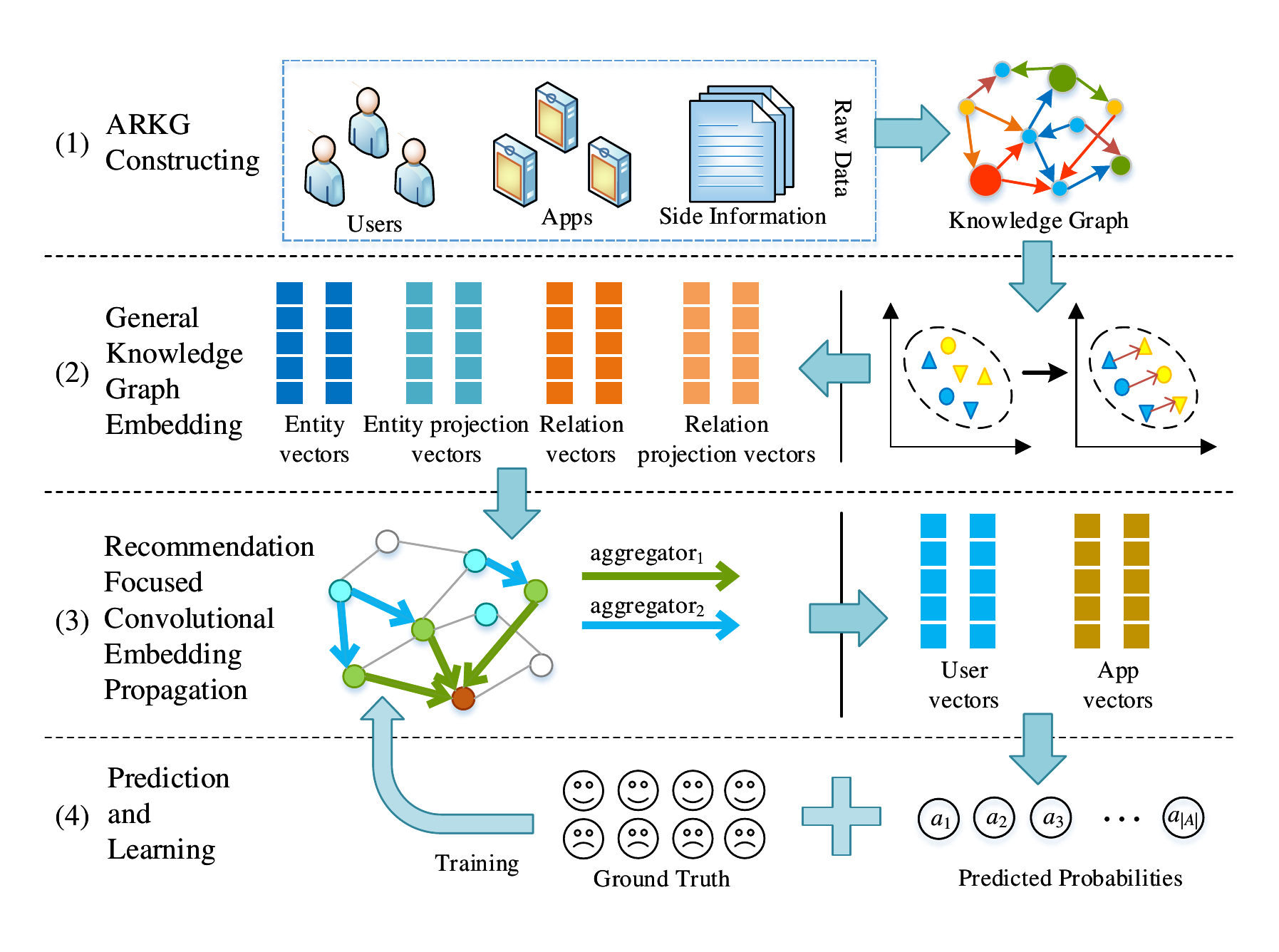}
\caption{Framework of the proposed Model KGEP} \label{framework}
\end{figure}

The framework of our app recommendation model KGEP was presented in Fig.~\ref{framework}, which consists of four main components:  1) ARKG constructing, which aims to construct an ARKG for app recommendation; 2) general KG embedding, which parameterizes each entity or relation as two vectors by preserving the semantic relatedness among the ARKG; 3) recommendation focused convolutional embedding propagation, which recursively propagates embeddings from a node’s tail neighbors to update its representation; 4) prediction and learning, which outputs the predicted matching scores by the final representations of users and apps, and learns the model parameters. We presented them in detail in the following subsections respectively. 

\vspace{-2mm}
\subsection{ARKG Construction}
ARKG construction mainly involves 2 sub-tasks, i.e., entity identification and relation extraction.

\vspace{-2mm}
\subsubsection{Entity Identification.}
Except \textit{Content-Topic} entities, the other kinds of entities listed in the above section can be explicitly identified from the side information of users and apps. So we just described the identification method of \textit{Content-Topic} entities here.

We adopted an LDA model to identify \textit{Content-Topic} entities. Its basic idea is that documents are represented as random mixtures over latent topics, where each topic is characterized by a distribution over words. The process of \textit{Content-Topic} entity identification can be summarized as follows.
\begin{enumerate}
\item \textbf{Text preprocessing and hyperparameter setting}. Taking the \textit{Readme} text of each app as a document, we can get a corpus, i.e., a collection of $|\mathcal{A}|$ documents. We first preprocessed the corpus, i.e., conducting tokenization, stop words removing, stemming, lemmatization and typo corrections. NLTK and Spacy packages of Python were used for these tasks. Then, we set the number of topics and other hyperparameters.
\item \textbf{LDA model learning}.  Given a preprocessed corpus of documents, we used variational EM algorithm to estimate parameters in LDA model. Then for each app $a$, let $\mathcal{Z} = {z_1, z_2, …, z_K}$ be the set of latent topics, we can obtain the parameters $\theta_a = {\theta_{a1}, \theta_{a2}, …, \theta_{aK}}$ of its Dirichlet distribution over $\mathcal{Z}$.
\item \textbf{\textit{Content-Topic} entity identifying}. For each app $a$, given its inferred Dirichlet distribution parameters $\theta_a = {\theta_{a1}, \theta_{a2}, …, \theta_{aK}}$ over $\mathcal{Z}$, it is defined that  app $a$ has \textit{Content-Topic} $z_k$ if and only if $!\exists \theta_{ai}|\theta_{ai} > \theta_{ak}, (1\leq i\leq K)$. Namely, an app has one and only \textit{Content-Topic} entity.
\end{enumerate}

\vspace{-2mm}
\subsubsection{Relation Extraction.} Due to space limitations, we only detailed the triplet extraction methods for relations \textit{CTSIMILAR} and \textit{USIMILAR}, which are more complex than the methods for other relations. 

The relation \textit{CTSIMILAR} represents the similarity among \textit{Content-Topic} entities in the ARKG. Let $\mathcal{W} = \{w_1, w_2, …, w_V\}$ be the set of words and $\mathcal{Z} = \{z_1, z_2, …, z_K\}$ be the set of latent topics, we can get the probability distribution $\Phi_k = \{\phi_{k1}, \phi_{k2}$, $…, \phi_{kV}\}$ of each topic $z_k (1 \leq k \leq K)$ over $\mathcal{W}$ using variational EM algorithm. Then, for any two \textit{Content-Topic} entities $z_i$ and $z_j$, we used Hellinger distance to measure their similarity.

\begin{equation}
    similarity\_ct(z_i, z_j)=\frac{1}{\sqrt{2}}\sqrt{\sum_{l=1}^V(\sqrt{\phi_{il}} - \sqrt{\phi_{jl}})^2}
\end{equation}

Given the \textit{Content-Topic-Similarity} threshold $cts (0<cts<1)$, there will be a relation $(z_i$, \textit{CTSIMILAR}, $z_j)$ if $similarity\_ct(z_i, z_j)\geq cts$. 

The relation \textit{USIMILAR} represents the similarity among \textit{User} entities. We used a user-app rating matrix to extract this kind of relations. For typical app recommendation scenario, a user can rate an app from “1 star” to “5 star”, where we transformed the rating grades from 0.2 to 1.0, and “none rating” to 0. Then, for each user $u_i (1\leq i\leq |\mathcal{U}|)$, we can get a rating vector $\mathbi{r}_i = \{r_{i1}, r_{i2}, …, r_{i|\mathcal{A}|}\}$. The similarity between any two users $u_i$ and $u_j$ are modeled as their Tanimoto coefficient.

\begin{equation}
\begin{aligned}
    similarity\_u(u_i, u_j)
    &= \frac{\mathbi{r}_i\cdot\mathbi{r}_j}{\|\mathbi{r}_i\|^2 + \|\mathbi{r}_j\|^2 - \mathbi{r}_i\cdot\mathbi{r}_j}\\
    &= \frac{\sum_{l=1}^{|\mathcal{A}|}r_{il}r_{jl}}{\sum_{l=1}^{|\mathcal{A}|}r_{il}^2 + \sum_{l=1}^{|\mathcal{A}|}r_{jl}^2 - \sum_{l=1}^{|\mathcal{A}|}r_{il}r_{jl}}
\end{aligned}
\end{equation}

Given the \textit{User-Similarity} threshold $us (0<us<1)$, we can extract a triplet $(u_i, USIMILAR, u_j)$ if \textit{similarity\_u} $(u_i, u_j)\geq us$ for $u_i$ and $u_j$.

\subsection{General KG Embedding}
General KG embedding was then performed on the constructed ARKG to embed its entities and relations into continuous vector spaces, while preserving its inherent structure. For the ARKG mainly consists of N-to-1 and N to N relations, we employed TransD\cite{TransD}, which is suitable for dealing with such complex relations and at the same time has relatively high efficiency, to embed the ARKG. 

To be more specific, for each triplet $(h, r, t)$ in the ARKG, it learns two vectors for the head entity $h$, tail entity $t$ and relation $r$ respectively, denoted as $\mathbf{h}$, $\mathbf{h}_p$, $\mathbf{t}$, $\mathbf{t}_p$, $\mathbf{r}$ and $\mathbf{r}_p$, where $\mathbf{h}, \mathbf{h}_p, \mathbf{t}, \mathbf{t}_p \in \mathbb{R}^m$ and $\mathbf{r}, \mathbf{r}_p \in \mathbb{R}^n$. We set the hyperparameters $m=n=d$ for the convenience of the learned vectors' application in recommendation. The prior vectors $\mathbf{h}$, $\mathbf{t}$ and $\mathbf{r}$ represent the meaning of the entities and relation. The other ones (i.e., $\mathbf{h}_p$, $\mathbf{t}_p$, $\mathbf{r}_p$) are called projection vectors representing the way that how to project the entity embeddings $\mathbf{h}$ and $\mathbf{t}$ into the relation vector $\mathbf{r}$'s space. Specifically, they were used to construct mapping matrices, which are defined as follows.
\begin{equation}
\begin{array}{c}
\mathbf{M}_{rh} = \mathbf{r}_p\mathbf{h}_p^\top + \mathbf{I}^{d\times d} \vspace{1ex} \\
\mathbf{M}_{rt} = \mathbf{r}_p\mathbf{t}_p^\top + \mathbf{I}^{d\times d}
\end{array}
\end{equation}
where $\mathbf{M}_{rh}$, $\mathbf{M}_{rh} \in \mathbb{R}^{d\times d}$ are mapping matrices, and $\mathbf{I}$ denotes the identity matrix of size $d\times d$. With the mapping matrices, the projected vectors of $\mathbf{h}$ and $\mathbf{t}$ are defined as follows.
\begin{equation}
\mathbf{h}_\perp = \mathbf{M}_{rh}\mathbf{h}, \hspace{2ex} \mathbf{t}_\perp = \mathbf{M}_{rt}\mathbf{h} 
\end{equation}
To learn embeddings of each entity and relation by optimizing the translation principle $\mathbf{h}_\perp + \mathbf{r} \approx \mathbf{t}_\perp$, the plausibility score (aka energy score) of a given triplet $(h, r, t)$ was formulated as follows.
\begin{equation}
g(h,r,t) = -\| \mathbf{h}_\perp +\mathbf{r}-\mathbf{t}_\perp\|_2^2
\end{equation}
where a higher score of $g(h,r,t)$ suggests that the triplet is more likely to be true, and vice versa.

The training of TransD uses the following margin-based ranking loss to encourage discrimination between golden triplets and incorrect ones.

\begin{equation}
\mathcal{L}_{K\!G}=\sum_{(h,r,t)\in S}\sum_{(h',r,t')\in S'}\max \big(0,\gamma+g(h',r,t')-g(h,r,t) \big)
\end{equation}

where $\max(x,y)$ aims to get the maximum between $x$ and $y$, $\gamma$ is the margin, $S=\{(h,r,t)\}$ is the set of golden triples contained in the ARKG. Corrupting each golden triplet $(h,r,t) \in S$ by replacing the head entity or the tail entity, the set of negative triples $S'=\{(h',r,t')\}$ can be generated. The process of minimizing the above objective was carried out with stochastic gradient descent (SGD) in mini-batch mode.

This component embeds the entities and relations on the granularity of triples. After getting its outputs, we can use them to make app recommendation directly and roughly by equation (5).  

\subsection{Convolutional Embedding Propagation}
\begin{figure}
\includegraphics[width=\textwidth]{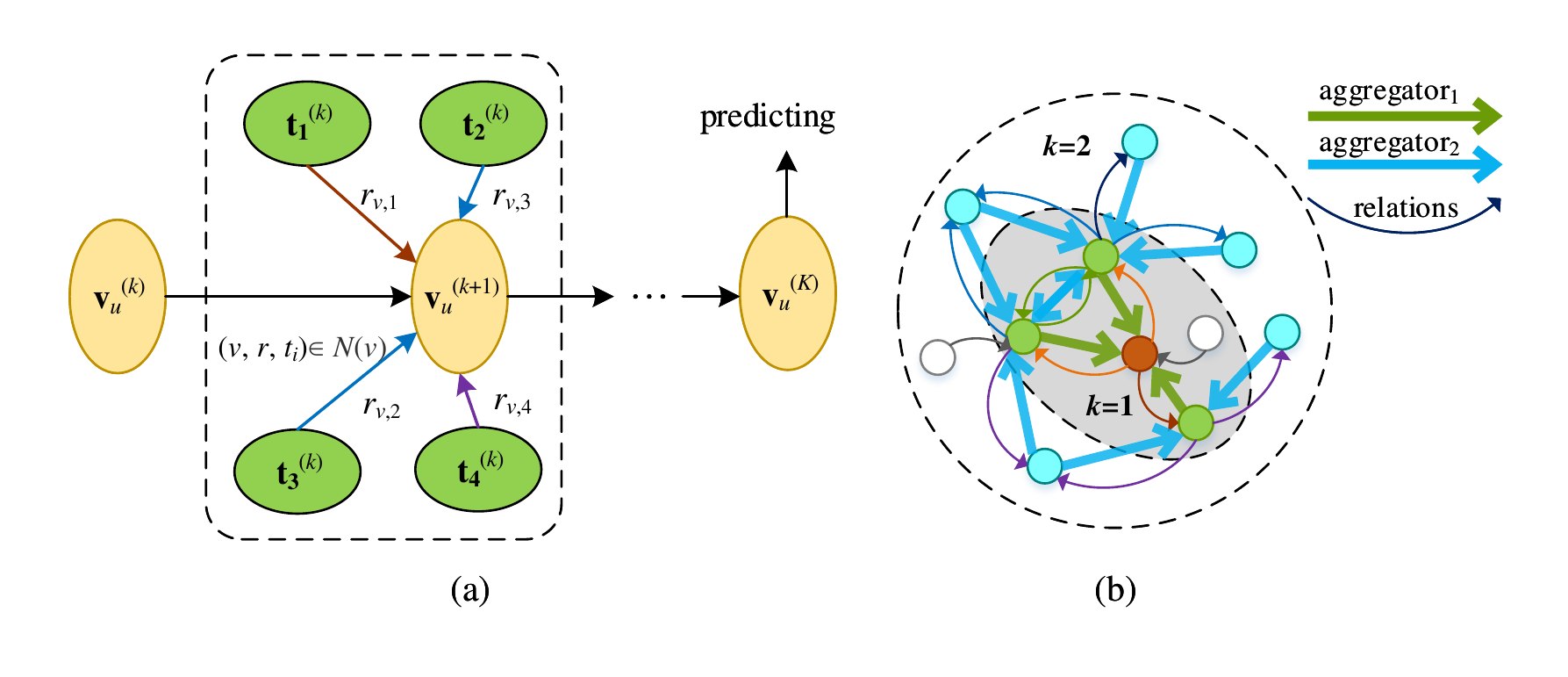}
\caption{Illustration of our convolutional embedding propagation approach. (a) an example of how an entity aggregates information from its neighbours. (b) an example of how the model propagates information between two layers.} \label{convolution}
\end{figure}

Next, focused on app recommendation, we built upon the architecture of graph convolution network to further capture both high-order structure and semantic information in the ARKG to make more precise recommendation. Here we started by describing a single layer, and then discussed how to generalize it to multiple layers. As illustrated in Fig.~\ref{convolution} (a), one single layer mainly involves 2 steps: 1) for each entity, aggregating information from its neighbors to form its neighbors' aggregated vector; 2) integrating with its own current latent vector to update its embedding for the next layer.

\vspace{-2mm}
\subsubsection{Aggregating Information from Neighbours.} 
In the real ARKG, head entities are causally determined by tail entities (e.g. the apps should be profiled by their attributes, and users' preference should be influenced by apps). So information is aggregated from tail entities to head entities in our model. In addition, to characterize both semantic information of the ARKG and users’ personalized interests in relations, neighbours are weighted dependent on the connecting relation and specific user while calculating the neighbours' aggregated vector for a given entity. Specifically, given a user $u$ and a node $v$ in the ARKG ($\mathcal{G}$), we use  $\mathcal{N}_v=\{(h,r,t)|(h=v)\wedge(h,r,t)\in\mathcal{G}\}$ to denote the set of triplets where $v$ is the head entity. Then the neighbours' aggregated vector of $v$ specific to $u$ is computed as follows. 
\begin{equation}
\mathbf{v}^{\mathcal{N}_v}_u=\sum_{(h,r,t)\in \mathcal{N}_v}w_u^r\mathbf{t}
\end{equation}
where $\mathbf{t}\in \mathbb{R}^d$ is the vector of tail entity $t$, and $w_u^r$ is the weight between user $u$ and relation $r$, which characterizes the importance of relation $r$ to user $u$ and be computed as follows.
\begin{equation}
w_u^r = \frac{\exp(\pi(\mathbf{u},\mathbf{r}))}{\sum_{(h,r,t)\in \mathcal{N}_v}\exp(\pi(\mathbf{u},\mathbf{r}))}
\end{equation}
where $\mathbf{u}\in \mathbb{R}^d$ and $\mathbf{r}\in \mathbb{R}^d$ are the embeddings of user $u$ and relation $r$. $\pi: \mathbb{R}^d \times \mathbb{R}^d \rightarrow \mathbb{R}^d$ is a weight score function (e.g., we adopted inner product in this paper). 

Generally, $\mathbf{v}^{\mathcal{N}_v}_u$ not only characterizes the local proximity structure of node $v$, but also exploits the personalized interests of user $u$ in relations.

\vspace{-2mm}
\subsubsection{Updating Embeddings for the Next Layer.}
To update the embedding of each node $v$ as its representation in the next layer, we concatenated its current representation $\mathbf{v}_u$ with its neighbours' aggregated vector $\mathbf{v}^{\mathcal{N}_v}_u$, and fed this concatenated vector through a fully connected layer with nonlinear activation function $\sigma$ to transform it to the new representation of $v$. It can be formulated as:
\begin{equation}
\mathbf{v}_u'=\sigma \big(\mathbf{W}\cdot(\mathbf{v}_u\|\mathbf{v}^{\mathcal{N}_v}_u)+\mathbf{b}\big)
\end{equation}
where $\mathbf{v}_u'$ (i.e., the output of this layer) is the new representation of node $v$ specific to user $u$, and $\mathbf{W}$ and $\mathbf{b}$ are transformation weight and bias, respectively. ``$\|$'' denotes the concatenation operation.

Note that not all entities are updated because some of them in the ARKG have no tail neighbours.

\vspace{-2mm}
\subsubsection{Information Propagating among Layers.} 

Through a single layer, we can capture 2-order entity connectivity, taking the general KG embedding as the 1-order connectivity. However, exploiting higher-order connectivity is of importance to perform high-quality recommendation. It is intuitive to propagate information between different layers to capture higher-order structural proximity among entities. As illustrated in Figure~\ref{convolution} (b),  given the brown entity, its embedding is updated by aggregating information from its neighbours (i.e., the green nodes), while the embeddings of the green nodes are updated by aggregating information from their neighbours (i.e., the blue ones). 

More formally, we stacked $K-1$ propagation layers and used equation (9) to  propagate embeddings along higher-order connectivity. For notational convenience, we denoted the representation of node $v$ specific to user $u$ at depth $k-1$ as $\mathbf{v}_u^{(k)}$. Generally speaking, $\mathbf{v}_u^{(k)}$ is a mixture of initial representations of node $v$ and its neighbors up to $k$ hops away. 

\subsection{Model Prediction and Learning}

After performing $K-1$ layers, we obtained the final representation $\mathbf{v}_u^{(K)}$ of node $v$ specific to user $u$, which characterizes $v$'s high-order entity dependencies up to $K$ hops and captures $u$'s potential long-distance interests. In addition, the outputs of the general KG embedding characterize the distance between head entity $h$ and tail entity $t$ in the space of relation $r$ for a triplet $(h,r,t)$. So for user $u$ and app $a$, we concatenated the representations of the two components into a single vector to do prediction as follows.
\begin{equation}
\mathbf{u}^* = (\mathbf{u}_\perp + \mathbf{r}_{I\!N\!T\!E\!R\!A\!C\!T}) \| \mathbf{u}^{(K)}, \hspace{5mm}
\mathbf{a}_u^* = \mathbf{a}_\perp \| \mathbf{a}_u^{(K)}
\end{equation}
where $\mathbf{r}_{I\!N\!T\!E\!R\!A\!C\!T}$ is the vector of relation ``\textit{INTERACT}'', $\mathbf{u}_\perp$ and $\mathbf{a}_\perp$ are the vectors of user $u$ and app $a$ in the $\mathbf{r}_{I\!N\!T\!E\!R\!A\!C\!T}$ space. They are all the outputs of the general KG embedding component, while $\mathbf{u}^{(K)}$ and $\mathbf{a}_u^{(K)}$ are the final outputs of the convolutional embedding propagating component.

Finally, we computed the inner product of user and app representations, so as to predict their matching score:
\begin{equation}
\hat{y}_{u,a} = \mathbf{u}^{*\top} \mathbf{a}_u^*
\end{equation}

To learn the parameters of our app recommendation model, we adopted negative sampling strategy, and the objective function was defined as binary cross-entropy loss with $L2$ norm regularization:
\begin{equation}
\mathcal{L}_{C\!E\!P}=\sum_{u\in\mathcal{U}}\sum_{v\in {T\!r\!n}^u}\big(-\log \hat{y}_{u,v} + \sum_{i\in {N\!e\!g}_v^u}-\log(1-\hat{y}_{u,i})\big)+\lambda\|\theta\|_2^2
\end{equation}
where $T\!r\!n^u=\{v|y_{u,v}=1\}$ is the set of user $u$'s all training instances. For each training instance $(u,v)$, we randomly sampled $x$ negative apps, denoted as $N\!e\!g_v^u$. $\lambda$ is coefficient for the regularization and $\theta$ denotes all model parameters. The model was trained via Adam optimizer.

\section{Empirical Study}
In this section, we compared our approach with several state-of-the-art recommendation methods using real-world app usage data and studied the impact of parameters on the performance of our model.

\subsection{Dataset Description}
We collected our dataset from Google Play. We crawled each app relevant metadata that the ARKG constructing needs. To bypass the cold start, we first omited apps with less than 10 users and then excluded users with less than 10 apps. After this preprocessing step, our dataset contains 12802 users, 4539 apps, and 198077 rating observations. The user app rating matrix has a sparsity as high as 0.341\%.

\subsection{Comparing Methods}
To evaluate the performance of the proposed model, we compared it with the following representative baselines.
\begin{enumerate}
\item \textbf{UserCF}: A user-user similarity matrix can be obtained while extracting \textit{USIMILAR} relations for the ARKG. Then, we used the classic user-based collaborative filtering as a baseline. 
\item\textbf{BPR-MF}\cite{BPR}: The Bayesian Personalized Ranking based matrix factorization, is a representative algorithm designed for implicit feedback, adopting a pairwise ranking loss to optimize the latent factor models.
\item\textbf{FISM}\cite{FISM}: This is representative item-based collaborative filtering Top-N recommendation model, used to verify the effectiveness of our recommendation model.
\item\textbf{NFM}\cite{NFM}: This is a state-of-the-art factorization model, which seamlessly combines FM and neural network in modelling feature interactions. 
\item\textbf{TransDR}: This is a simplified translation-based recommendation version of our approach, which takes the representation learned by TransD as inputs of a one hidden layer neural network to make recommendation directly.
\end{enumerate}

\subsection{Experiment Setup}
We divided the preprocessed dataset into three subsets: training, validation, and test. For every user, we randomly selected 70\%, 10\% and 20\%  interacted apps into the training set, the validation set and the test set respectively.

For ARKG construction, we set the number of \textit{Content-Topic} entities to 50, the \textit{Content-Topic-Similarity} threshold $cts$ to 0.9, and the \textit{User-Similarity} threshold $us$ to 0.98, and then we extracted 406044 triplets for the ARKG. 

We implemented our KGEP model in Tensorflow. The hyper-parameters were optimized on the validation set, which are listed as follows. The embedding size is 16, the number of propagation layers is 1, dropout is 0, epoch is 80, and learning rate is 0.02. All the experiment results of our model are corresponding to the above hyper-parameter values, except a specific hyper-parameter may vary while preserving the values of the other  hyper-parameters when we analyzed our model sensitivity on the given hyper-parameter. For all the baselines, we set respective optimal parameters either according to corresponding references or based on our experiment results. We adopted learner \textit{Adam} for the models: BPR-MF, FISM, NFM and KGEP, and adopt learner \textit{SDG} for TransDR. 

We adopted three widely used metrics for performance evaluation: $Recall@K$, $Precision@K$ and mean average precision ($M\!A\!P@K$), where $K$ indicates recommending top $K$ ranked apps. For all the metrics, the larger the value, the better the performance.

\subsection{Performance Comparison with Baseline Methods}
The performance comparison results are presented in Table~\ref{comparison}. We had the following observations:
(1) KGEP consistently yields the best performance on all the metrics and $K$ values. In detail, KGEP improves much more over the strongest baselines on the metric MAP than the other 2 metrics, and when $K$ is smaller; 
(2) BPR-MF achieves better performance than the other baselines in most cases;
(3) TransDR sometimes achieves better performance than all the baselines, indicating that just general KG embedding has efficacy to some extent to make app recommendation.

\begin{table}[]
\caption{Performance comparison on the Google Play dataset. The best results are starred, and the second-best results are listed in bold.}\label{comparison}
\small
\begin{tabular}{c|p{2.2cm}|p{1.35cm}<{\centering}|p{1.35cm}<{\centering}|p{1.35cm}<{\centering}|p{1.35cm}<{\centering}|p{1.35cm}<{\centering}|p{1.35cm}<{\centering}}
\hline
top-K                & Metrics          & UserCF & BRP-MF         & FISM  & NFM   & TransDR        & KGEP            \\
\hline
                     & Precision   (\%) & 0.155  & \textbf{0.458} & 0.209 & 0.301 & 0.315          & \textbf{1.000}* \\
                     & Recall   (\%)    & 0.254  & \textbf{1.159} & 0.521 & 0.763 & 0.567          & \textbf{2.461}* \\
\multirow{-3}{*}{10} & MAP   (\%)       & 0.361  & \textbf{1.309} & 0.656 & 0.859 & 0.871          & \textbf{3.853}* \\
\hline
                     & Precision   (\%) & 0.144  & \textbf{0.440} & 0.199 & 0.277 & 0.390          & \textbf{0.600}* \\
                     & Recall   (\%)    & 0.341  & \textbf{2.187} & 0.997 & 1.352 & 1.151          & \textbf{3.061}* \\
\multirow{-3}{*}{20} & MAP   (\%)       & 0.398  & \textbf{1.516} & 0.769 & 1.001 & 1.285          & \textbf{3.996}* \\
\hline
                     & Precision   (\%) & 0.145  & 0.426          & 0.200 & 0.265 & \textbf{0.430} & \textbf{0.567}* \\
                     & Recall   (\%)    & 0.452  & \textbf{3.183} & 1.491 & 1.864 & 1.749          & \textbf{4.256}* \\
\multirow{-3}{*}{30} & MAP   (\%)       & 0.428  & \textbf{1.616} & 0.821 & 1.069 & 1.510          & \textbf{4.177}* \\
\hline
                     & Precision   (\%) & 0.145  & 0.413          & 0.191 & 0.258 & \textbf{0.461} & \textbf{0.475}* \\
                     & Recall   (\%)    & 0.550  & \textbf{4.034} & 1.864 & 2.450 & 2.385          & \textbf{4.839}* \\
\multirow{-3}{*}{40} & MAP   (\%)       & 0.450  & \textbf{1.667} & 0.848 & 1.103 & 1.599          & \textbf{4.232}* \\
\hline
\end{tabular}
\end{table}

\subsection{Model Analysis and Discussion}
\begin{figure}
\includegraphics[width=\textwidth]{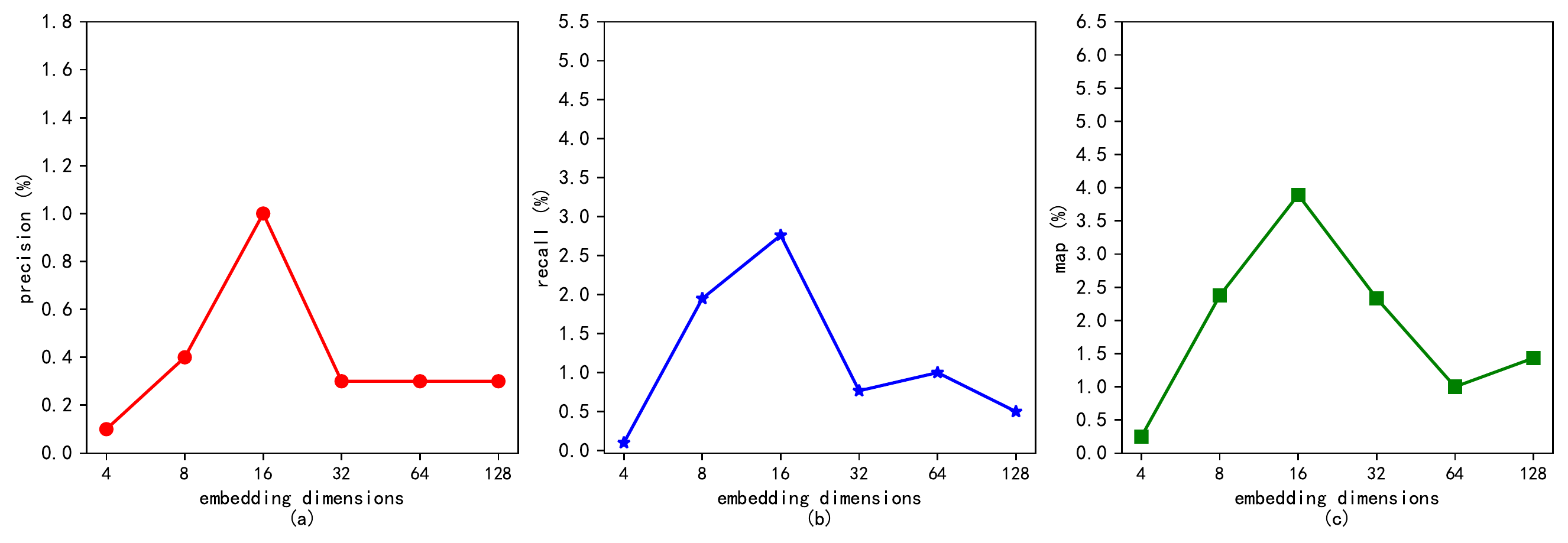}
\caption{Effect of embedding size} \label{embeddingsize}
\end{figure}

To get deep insights on the proposed model KGEP, we investigated its sensitivity on some core hyper-parameters. Fig.~\ref{embeddingsize} illustrates the effect of embedding size. Due to the computational cost, we can not train TransD model after the embedding size is larger than 16. So, we used xavier initializer to initialize the propagation embeddings after the dimensionality is larger than 16, and concatenated the 16-dimensional embeddings of TransD to make recommendation. From Fig.~\ref{embeddingsize}, we can see, our model KGEP can achieve the best performance when the embedding size is set to 16.  

\vspace{-3mm}
\begin{figure}
\includegraphics[width=\textwidth]{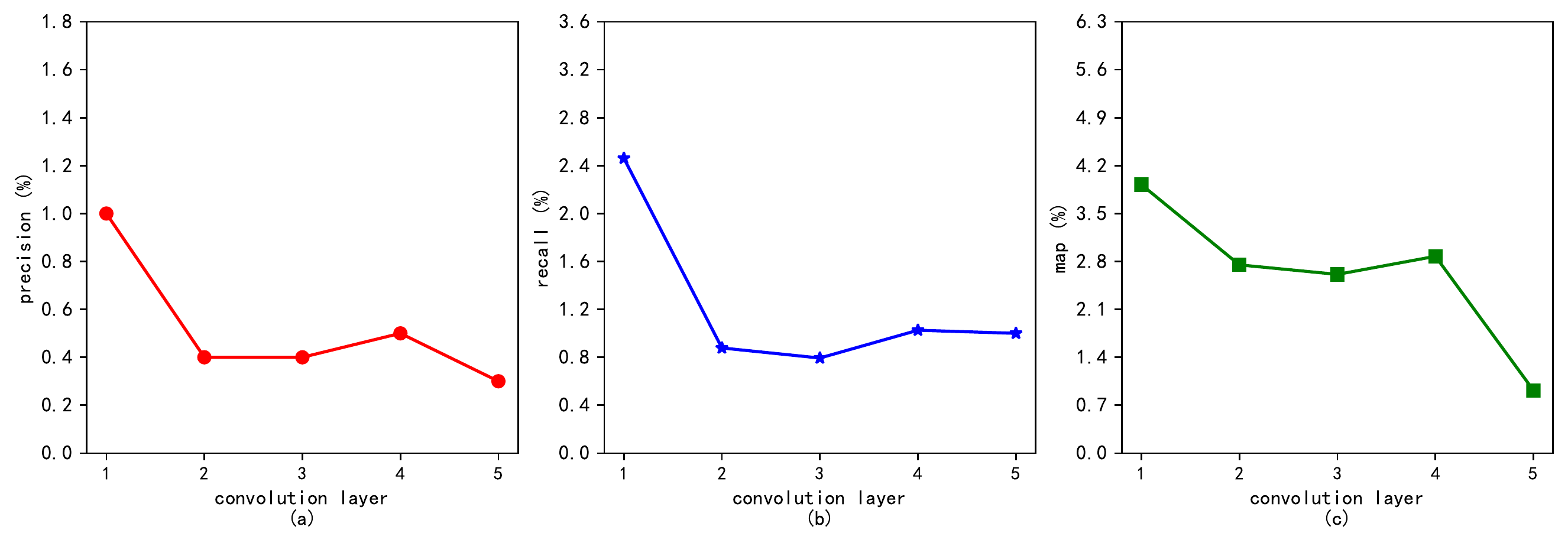}
\caption{Effect of embedding propagation layer numbers} \label{layernumber}
\end{figure}

Fig.~\ref{layernumber} shows the influence of layer numbers. It illustrates that our model achieves the best performance just with one embedding propagation layer on the basis of general ARKG embeddings. We also conducted the experiments to analyze the effects of dropout and learning rate. Due to space limitations, the corresponding figures are not presented any more. The results are that the performance of KGEP would be the best when the learning rate equals to 0.02 among \{0.0001, 0.0005, 0.001, 0.05, 0.02, 0.1, 0.5\}, and would be better when the dropout equals to 0 or 0.2 than other values \{0.1, 0.3, 0.4, 0.5, 0.6, 0.7, 0.8, 0.9\}. 

\section{Related Work}

\subsubsection{Knowledge Graph based Recommendation.}
Recommender systems are now indispensable in many Web applications, such as App stores. The matrix factorization algorithm BPR-MF\cite{BPR}, the item-based collaborative filtering algorithm FISM\cite{FISM}, and the factorization model algorithm NFM\cite{NFM} are arguably the most representative among the large number of recommendation algorithms developed. Recently, KG, as one of the most effective data modelling techniques, has been spotlighted in recommender systems. In general, existing KG-aware recommendation can be classified into three categories. The first category is embedding-based methods, such as CKE\cite{CKE}, DKN\cite{DKN}, which preprocess a KG with knowledge graph embedding algorithms and incorporates the learned entity embeddings into a recommendation framework. However, these methods are usually more suitable for in-graph applications such as link prediction than for recommendation. The second category is path-based methods, such as PER\cite{PER}, KPRN\cite{KPRN}, which explore the various patterns of connections among items in KG to provide additional guidance for recommendations. However, they rely heavily on meta-paths, which is hard to optimize in practice, so that has a large impact on the final recommendation performance. The third category is embedding propagation methods, such as RippleNet\cite{RippleNet}, KGAT\cite{KGAT}, KGCN\cite{KGCN}, which combine embedding-based and path-based methods in KG-aware recommendation, so as to address the limitations of the above two categories. 

Different from the above KG-aware recommendation models, we leverage the general embeddings and the propagated embeddings simultaneously to make app recommendation.

\vspace{-2 mm}
\subsubsection{Mobile App Recommendation.}
Mobile app recommendation has attracted much attention these days. By focusing on different kinds of side information, researchers proposed the following representative app recommendation approaches. Focusing on the privilege data of apps, The studies\cite{ARLiu,ARHuang,ARYin} mainly considered privacy leak and security risk issues to perform personalized app recommendations. Focusing on geographical information of users, Zhu et al.\cite{ARZhu} proposed a novel location-based probabilistic factor analysis mechanism to help people get an appropriate mobile app. Focusing on version information of apps, Cao et al.\cite{ARCao} proposed a novel version-sensitive mobile app recommendation framework by jointly exploring the version progression and dual-heterogeneous data. Focusing on app usage patterns of users, Xu et al. \cite{ARXu} proposed a neural network based approach to leverage the predictive power of app usage context patterns to do effective app recommendation. Focusing on category information of apps, Guo et al. \cite{ARGuo} proposed an app recommendation model based on deep factorization machine, which can make use of categorical and textual information of apps. Further considering the interactions of categories and other side information of apps, Liang et al.\cite{ARLiang} utilized a tensor-based framework to effectively integrate app category information and multi-view features of users and apps to do context-aware app recommendation. Focusing on the complex semantics among different kinds of side information, Xie et al. \cite{ARXie} exploited weighted meta-graph and heterogeneous information network for mobile app recommendation, mainly considering user review information. However, it is not an end-to-end method. Meta-graphs are hard to be designed optimally, which will further influence the efficacy of recommendation.

Differed from the above state-of-the-art app recommendation methods, we proposed an end-to-end framework and leveraged KG to recommend apps for users. It can model complex semantics among diverse side information more explicitly to make better recommendation.

\section{Conclusion and Future Work}
This paper proposed a novel KG based mobile app recommendation approach. We first designed a KG construction method to organize rich side information of users and apps, then adopted a translation based KG embedding method to capture the semantics of side information related to first-order structure of the constructed KG, and proposed a  convolutional embedding propagation model to capture the semantics related to high-order structure of the KG. By incorporating KG into app recommendation, our approach can effectively model and take full advantage of rich side information to alleviate the sparsity issue and improve recommendation performance. The comparative experimental results show that our approach outperforms the competing recommendation methods in terms of precision, recall and MAP.
 
In the future, we will attempt to apply our model to other recommendation application scenarios, such as general Web service recommendation or  Web API recommendation for Mashups, to further validate it and find and improve its limitations. 

\vspace{-1mm}
%
%
%

\begin{thebibliography}{88}
\bibitem{GP}
Number of Android apps on Google Play, \url{https://www.appbrain.com/stats/number-of-android-apps}.Accessed 12
May 2020

\bibitem{ARLiu}
Liu, B., Kong, D., Cen, L., Gong, N. Z., Jin, H., Xiong, H.: Personalized mobile app recommendation: reconciling app functionality and user privacy preference. In: WSDM, pp. 315--324. ACM, New York (2015)

\bibitem{ARHuang}
Huang, K., Han, J., Chen, S., Feng, Z.: A skewness-based framework for mobile app permission recommendation and risk evaluation. In: Sheng, Q.Z., Stroulia, E., Tata, S., Bhiri, S. (eds.) ICSOC 2016. LNCS, vol. 9936, pp. 252–266. Springer, Cham (2016). \doi{https://doi.org/10.1007/978-3-319-46295-0 16}

\bibitem{ARYin}
Yin, H., Chen, L., Wang, W., Du, X., Nguyen, Q.V.H., Zhou, X.: Mobi-SAGE: a sparse additive generative model for mobile app recommendation. In: ICDE, pp. 75–78. IEEE, Piscataway (2017)

\bibitem{ARZhu}
Zhu, K., Zhang, L., Pattavina, A.: Learning geographical and mobility factors for mobile application recommendation. IEEE Intelligent Systems \textbf{32}(3), 36–44 (2017)

\bibitem{ARCao}
Cao, D., Nie, L., He, X., Wei, X., Shen, J., Wu, S., Chua, T. S.: Version-sensitive mobile app recommendation. Information Science \textbf{381}(1), 161–175 (2017)

\bibitem{ARXu}
Xu, Y., Zhu, Y., Shen, Y., Yu, J.: Leveraging app usage contexts for app recommendation: a neural approach. World Wide Web Journal \textbf{22}(6), 2721--2745 (2019)

\bibitem{ARGuo}
Guo, C., Xu, Y., Hou, X., Dong, N., Xu, J., Ye, Q.: Deep attentive factorization machine for app recommendation service. In: ICWS, pp. 134--138. IEEE, Piscataway (2019)

\bibitem{ARLiang}
Liang, T., He, L., Lu, C.T., Chen, L., Yu, P.S., Wu, J.: A broad learning approach for context-aware mobile application recommendation. In: ICDM, pp. 955–960. IEEE, Piscataway (2017)

\bibitem{ARXie}
Xie, F., Chen, L., Ye, Y., Liu, Y., Zheng, Z., Liu, X.: A weighted meta-graph based approach for mobile application recommendation on heterogeneous information networks. In: Pahl, C., Vukovic, M., Yin, J., Yu, Q. (eds.) ICSOC 2018, LNCS, vol. 11236, pp. 404--420. Springer, Cham (2018). \doi{https://doi.org/10.1007/978-3-030-03596-9\_29}

\bibitem{KG}
Ji, S., Pan, S., Cambria, E., Marttinen, P., Yu, P. S.: A survey on knowledge graphs: representation, acquisition and applications. arXiv preprint (2020). \url{https://arxiv.org/abs/2002.00388}

\bibitem{TransD}
Ji, G., He, S., Xu, L., Liu, K., Zhao, J.: Knowledge graph embedding via dynamic mapping matrix. In: ACL, pp. 687--696. The Association for Computer Linguistics, Stroudsburg (2015)

\bibitem{BPR}
Rendle, S., Freudenthaler, C., Gantner, Z., Schmidt-Thieme, L.: BPR: Bayesian personalized ranking from implicit feedback. In: UAI, pp. 452–461. AUAI, Corvallis, Oregon (2009)

\bibitem{FISM}
Kabbur, S., Ning, X., Karypis, G.: FISM: factored item similarity models for top-N recommender systems. In: SIGKDD, pp. 659--667. ACM, New York (2013)

\bibitem{NFM} 
He, X., Chua, T. S.: Neural Factorization Machines for Sparse Predictive Analytics. In: SIGIR, pp. 355--364. ACM, New York (2017)

\bibitem{CKE}
Zhang, F., Yuan, N. J., Lian, D., Xie, X., Ma, W. Y.: Collaborative knowledge base embedding for recommender systems. In: SIGKDD, pp. 353–362. ACM, New York (2016)

\bibitem{DKN}
Wang, H., Zhang, F., Xie, X., Guo, M.: DKN: Deep knowledge-aware network for news recommendation. In: WWW, pp. 1835–1844. ACM, New York (2018)

\bibitem{PER}
Yu, X., Ren, X., Sun, Y., Gu, Q., Sturt, B., Khandelwal, U., Norick, B., Han, J.: Personalized entity recommendation: a heterogeneous information network approach. In: WSDM, pp. 283–292. ACM, New York (2014)

\bibitem{KPRN}
Wang, X., Wang, D., Xu, C., He, X., Cao, Y., Chua, T. S.: Explainable reasoning over knowledge graphs for recommendation. In: AAAI, pp. 5329--5336. AAAI Press, Menlo Park (2019)

\bibitem{RippleNet}
Wang, H., Zhang, F., Wang, J., Zhao, M., Li, W., Xie, X., Guo, M.: RippleNet: propagating user preferences on the knowledge graph for recommender systems. In: CIKM, pp. 417--426. ACM, New York (2018)

\bibitem{KGAT}
Wang, X., He, X., Cao, Y., Liu, M., Chua, T. S.: KGAT: knowledge graph attention network for recommendation. In: SIGKDD, pp. 950--958. ACM, New York (2019)

\bibitem{KGCN}
Wang, H., Zhao, M., Xie, X., Li, W., Guo, M.: Knowledge graph convolutional networks for recommender systems. In: WWW, pp. 3307--3313. ACM, New York (2019)

\end{thebibliography}
%

\end{document}